\documentclass[bibyear]{aa} 

\usepackage{lscape}
\usepackage{float}

\usepackage{graphicx}
\usepackage{txfonts}
\begin{document} 

\title{WISE~J044232.92+322734.9: A star-forming galaxy at redshift 1.1 seen through a Galactic dust clump?}

   \author{O.~Miettinen}

   \institute{Academy of Finland, Hakaniemenranta 6, P.O. Box 131, FI-00531 Helsinki, Finland \\ \email{oskari.miettinen@aka.fi}}

   \date{Received ; accepted}

\authorrunning{Miettinen}
\titlerunning{J044232.92+322734.9}

\abstract{Physically unassociated background or foreground objects seen towards submillimetre sources are potential contaminants of both the studies of young stellar objects embedded in Galactic dust clumps and multiwavelength counterparts of submillimetre galaxies (SMGs).}{We aim to search for and characterise the properties of a potential extragalactic object seen in projection towards a Galactic dust clump.}{We employed the near-infrared (3.4~$\mu$m and 4.6~$\mu$m) and mid-infrared (12~$\mu$m and 22~$\mu$m) data from the \textit{Wide-field Infrared Survey Explorer (WISE)} and the submillimetre data from the \textit{Planck} satellite.}{We uncovered a source, namely 
the \textit{WISE} source J044232.92+322734.9 (hereafter J044232.92), which is detected in the W1--W3 bands of \textit{WISE}, but undetected at 22~$\mu$m (W4), and whose \textit{WISE} infrared (IR) colours suggest that it is a star-forming galaxy (SFG). This source is seen in projection towards the \textit{Planck}-detected dust clump PGCC~G169.20-8.96, which likely belongs to the Taurus-Auriga cloud complex, at a distance of 140~pc. We used the {\tt MAGPHYS+photo-}$z$ spectral energy distribution (SED) code to derive the photometric redshift and physical properties of J044232.92. The redshift was derived to be $z_{\rm phot}=1.132^{+0.280}_{-0.165}$, while, for example, the stellar mass, IR (8--1\,000~$\mu$m) luminosity, and star formation rate were derived to be $M_{\star}=4.6^{+4.7}_{-2.5}\times10^{11}$~M$_{\sun}$, $L_{\rm IR}=2.8^{+5.7}_{-1.5}\times10^{12}$~L$_{\sun}$, and ${\rm SFR}=191^{+580}_{-146}$~${\rm M}_{\sun}$~yr$^{-1}$ (or $281^{+569}_{-155}$~${\rm M}_{\sun}$~yr$^{-1}$ when estimated from the IR luminosity). The derived value of $L_{\rm IR}$ suggests that J044232.92 could be an ultraluminous IR galaxy, and we found that it is consistent with a main sequence SFG at a redshift of 1.132.}{The estimated physical properties of J044232.92 are comparable to those of SMGs, except that the derived stellar mass of J044232.92 appears somewhat higher (by a factor of 4--5) than the average stellar masses of SMGs. However, the stellar mass difference could just reflect the poorly sampled SED in the ultraviolet, optical, and near-IR regimes. Indeed, the SED of J044232.92 could not be well constrained using the currently available data (\textit{WISE} only), and hence the derived redshift of the source and its physical properties should be taken as preliminary estimates. Further observations, in particular high-resolution (sub-)millimetre and radio continuum imaging, are needed to better constrain the redshift and physical properties of J044232.92 and to see if the source really is a galaxy seen through a Galactic dust clump, in particular an SMG population member at $z\sim1.1$.}


   \maketitle
%

\section{Introduction}

Identifying the multiwavelength counterparts of an astrophysical object is an 
important step towards understanding the physical properties of the source. For example, 
if a Galactic molecular cloud core is found to be associated with a 
mid-infrared (IR) point source, the core is likely to harbour an embedded protostar 
(as opposed to a starless or prestellar core; see e.g. \cite{dunham2014} for a review). 
A corresponding example in extragalactic astronomy is the identification of multiwavelength counterparts of galaxies. 
For example, a reliable identification of the observed-frame optical, near-IR, and mid-IR counterparts of 
the most intense dusty star-forming galaxies (SFGs) in the observable Universe, the so-called submillimetre galaxies 
(SMGs; see \cite{casey2014}; \cite{hodge2020} for reviews), is a prerequisite to determine their 
photometric redshifts and physical properties through spectral energy distribution (SED) analysis (e.g. \cite{miettinen2017}; \cite{brisbin2017}).
 
A Galactic molecular cloud clump or core can have a background star (or stars) seen through the object's dusty medium (e.g. \cite{alves2001}). 
A similar chance alignment can also happen with foreground stars, but the stars are not reddened as in the case of background stars. In the case of SMGs, a background or a foreground galaxy can be incorrectly interpreted to be physically associated with the observed-frame submillimetre dust emission, especially if the projected angular separation between the two is on a sub-arcsecond scale. Such source contamination will then lead to a wrong redshift of the SMG (either too high or too low) and correspondingly incorrect physical properties, such as the stellar mass content and luminosity. 

In this paper, we present a potential case where an extragalactic object, namely an SFG, is projectively seen towards a Galactic molecular cloud clump. The target sources and data are described in Sect.~2, while the analysis and results are described in Sect.~3. The results are discussed in Sect.~4, and Sect.~5 summarises our results and main conclusions. 

Throughout this paper, we assume a $\Lambda$CDM (Lambda Cold Dark Matter) cosmology with the dark energy density $\Omega_\Lambda = 0.7$ and total matter density $\Omega_{\rm m} = 0.3$, while the Hubble constant is set to $H_0=70$~km~s$^{-1}$~Mpc$^{-1}$. We report the magnitudes in the Vega system, and notations such as $[3.4]$ refer to the Vega magnitude at a wavelength of 3.4~$\mu$m. We adopt a Chabrier (2003) Galactic-disc initial mass function (IMF). We define the spectral index, $\alpha$, as $S_{\nu}\propto \nu^{\alpha}$, where $S_{\nu}$ is the flux density at frequency $\nu$.

\section{Target sources and observations}

To search for a potential example case where an external galaxy is seen through a Galactic dust clump, we started from the whole-sky observations made with the \textit{Wide-field Infrared Survey Explorer} (\textit{WISE}; \cite{wright2010}) at 3.4, 4.6, 12, and 22~$\mu$m. As discussed by Koenig et al. (2012), SFGs exhibit an elevation in the emission from polycyclic aromatic hydrocarbons, and this phenomenon
can be seen in the \textit{WISE} $[4.6]-[12]$ IR colour of the galaxy. On the basis of the earlier \textit{Spitzer} IR colour criteria (e.g. \cite{gutermuth2008}), Koenig et al. (2012) determined the following \textit{WISE} IR colour criteria to identify SFGs (see Appendix A.1 therein):

\begin{equation}
\label{eqn:crit}
\begin{cases} [3.4]-[4.6]<0.46\times([4.6]-[12]-1.7) \\ [3.4]-[4.6]>-0.06\times([4.6]-[12]-4.67) \\ [3.4]-[4.6]<-1.0\times([4.6]-[12]-5.1) 
\\ [3.4]-[4.6]>0.48\times([4.6]-[12]-4.1) \\ [4.6]>12 \\ [4.6]-[12]>2.3 \end{cases}
\end{equation}
We note that there are also \textit{WISE} IR colour criteria to identify potential active galactic nuclei (AGN; \cite{koenig2012}; Appendix A.1 therein), but AGN are not considered in the present study.

We searched for sources from the AllWISE catalogue\footnote{{\tt https://irsa.ipac.caltech.edu/data/download/wise-\\allwise/}} that fulfil all the criteria in Eq.~(\ref{eqn:crit}). We note that the source needs to be detected in the \textit{WISE} bands W1--W3 (3.6, 4.6, and 12~$\mu$m) so that the inequalities in Eq.~(\ref{eqn:crit}) are well defined. To avoid very red sources that are strong in the 22~$\mu$m (W4) band of \textit{WISE}, for example the young stellar objects (YSOs) embedded in Galactic molecular clouds, the search was, somewhat arbitrarily, limited to sources with $[22]>5$. For comparison, about 79\% of the known YSOs in the Taurus-Auriga region in the Rebull et al. (2011) catalogue have 22~$\mu$m magnitudes of $[22]\leq5$. As an additional constraint, the \textit{WISE} source was required to be unaffected by known artefacts, that is the {\tt ccf} flag in the AllWISE catalogue was set to 0000. 

To search for potential foreground dust clumps, we employed the \textit{Planck}\footnote{\textit{Planck} (\url{http://www.esa.int/Planck}) is a project of the European Space Agency (ESA) with instruments provided by two scientific consortia funded by ESA member states (in particular the lead countries France and Italy), with contributions from NASA (USA) and telescope reflectors provided by a collaboration between ESA and a scientific consortium led and funded by Denmark.} Catalogue of Galactic Cold Clumps (PGCC; \cite{planck2016}). The PGCC catalogue contains 13\,188 Galactic sources spread across the whole sky (13\,242 sources in total, of which 54 are in the Small and Large Magellanic Clouds), including those at high Galactic latitudes, and hence it is well suited for our purpose because there is a higher probability of finding a background extragalactic source at high Galactic latitudes than close to the Galactic plane. For example, 27\% of the PGCC sources lie at Galactic latitudes between $b=5\degr$ and $b=40\degr$, and 31\% are between $b=-5\degr$ and $b=-40\degr$ (cf.~Fig.~3 in \cite{planck2016}). Another important advantage of PGCC is that about 42\% of its sources (5\,574) have reliable distance estimates available. This allowed us to confirm that the dust clump belongs to the Galaxy, and a known distance also allows the determination of the physical properties of the clump, such as its projected physical size and mass, which are also available in the PGCC catalogue. The right ascension distribution of the sources in the PGCC catalogue is double-peaked so that 44\% of the sources lie at $\alpha_{2000.0}=40\degr-150\degr$ and 40\% of the sources are at $\alpha_{2000.0}=230\degr-360\degr$. In the present study, we focussed on the former right ascension interval. 

It should be noted that the AllWISE catalogue is extremely large and contains data for almost $7.5\times10^8$ sources, and the full catalogue file size is 1.225~terabytes. For this reason, our search made use of only a fraction of the AllWISE catalogue. In practice, our source search procedure was excuted as follows. We first downloaded a sample of million sources from the AllWISE catalogue (1.3 per mille of the full catalogue) that have the magnitudes $[4.6]>12$ (the fifth inequality in Eq.~(\ref{eqn:crit})) and $[22]>5$, are free of known artefacts, and lie at $\alpha_{2000.0}=40\degr-150\degr$. After applying the remaining colour criteria in Eq.~(\ref{eqn:crit}), we were left with 318\,958 \textit{WISE} sources. The latter catalogue was cross-matched with the PGCC catalogue using a search radius of $10\arcsec$, which is a somewhat arbitrary choice and only 3.5\%--3.9\% of the \textit{Planck} beam size (full width at half maximum, FWHM) at 350~$\mu$m--850~$\mu$m (see Table~1 in \cite{planck2016}). Only one case was found, namely WISE~J044232.92+322734.9 (hereafter J044232.92), which is seen towards the \textit{Planck} source PGCC~G169.20-8.96 (hereafter G169.20-8.96), and where the projected angular separation between the two is $4\farcs15$.

\begin{figure}[!htb]
\centering
\resizebox{\hsize}{!}{\includegraphics{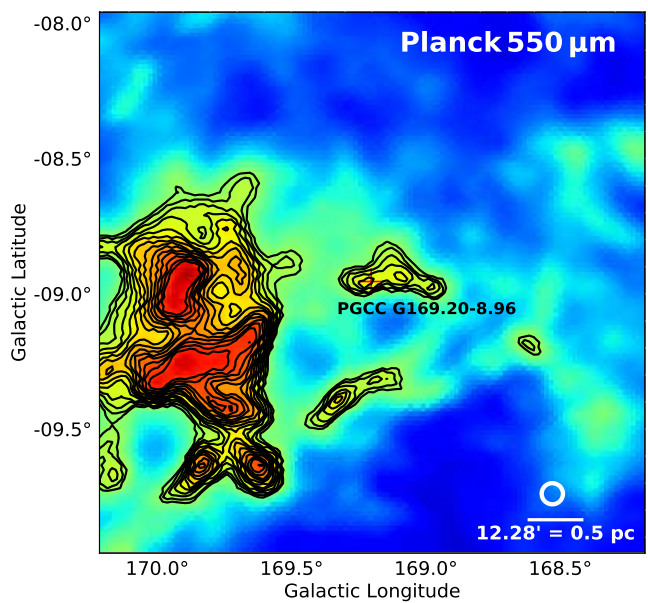}}
\caption{\textit{Planck} 550~$\mu$m image towards G169.20-8.96. The image is centred on the PGCC catalogue position of G169.20-8.96 and is $2\degr \times 2\degr$ in size. The colour scale is shown with linear scaling. The contour levels start at 12.8~MJy~sr$^{-1}$ and progress in steps of 0.4~MJy~sr$^{-1}$. The red plus sign indicates the position of the \textit{WISE} source J044232.92. The white circle shows the \textit{Planck} beam FWHM at 550~$\mu$m ($4\farcm682$). The scale bar corresponds to 0.5~pc at the distance of the dust clump 
G169.20-8.96 ($d=140$~pc).}
\label{figure:planck}
\end{figure}

\begin{figure*}[!htb]
\begin{center}
\includegraphics[width=0.268\textwidth]{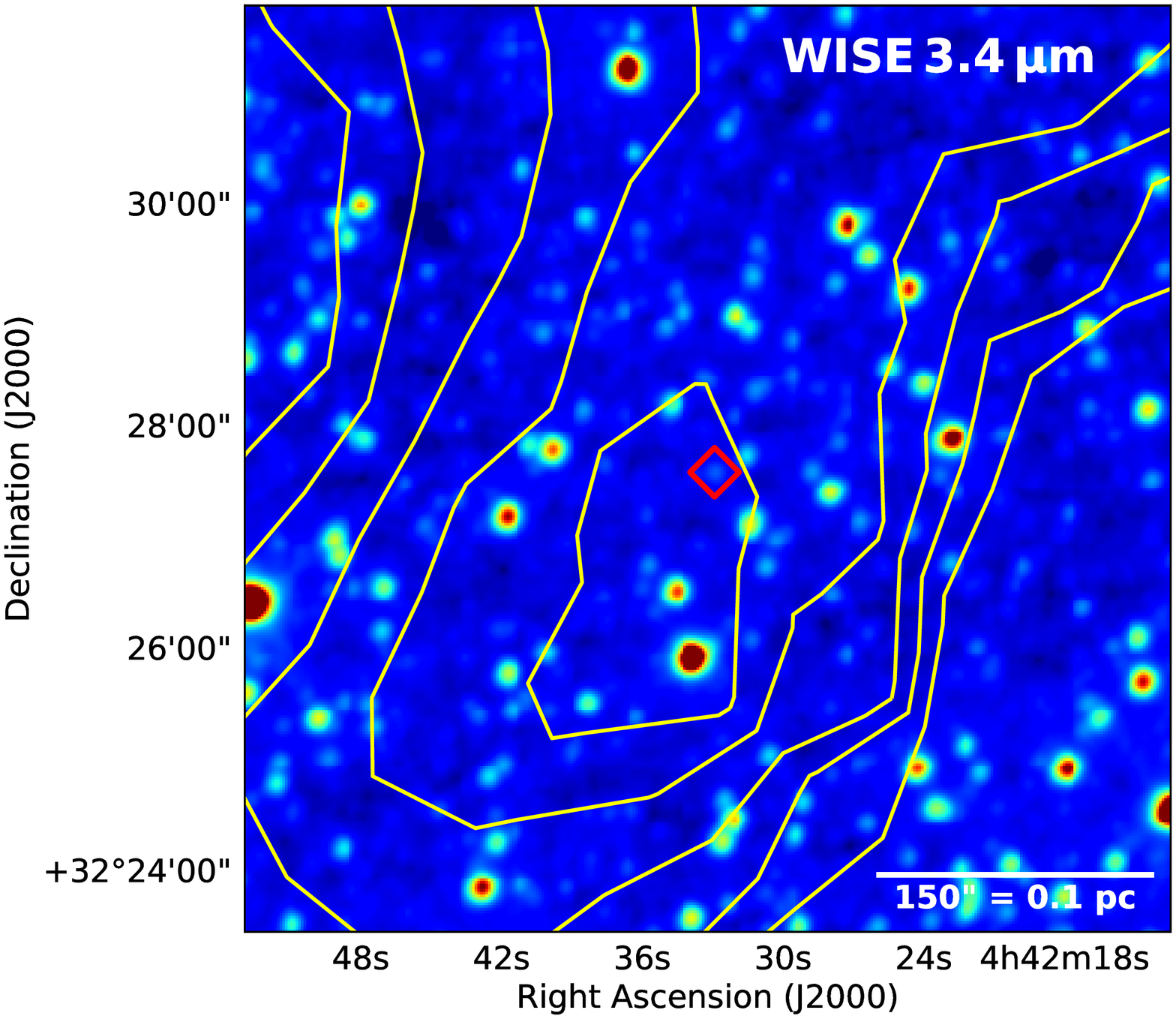}
\includegraphics[width=0.23\textwidth]{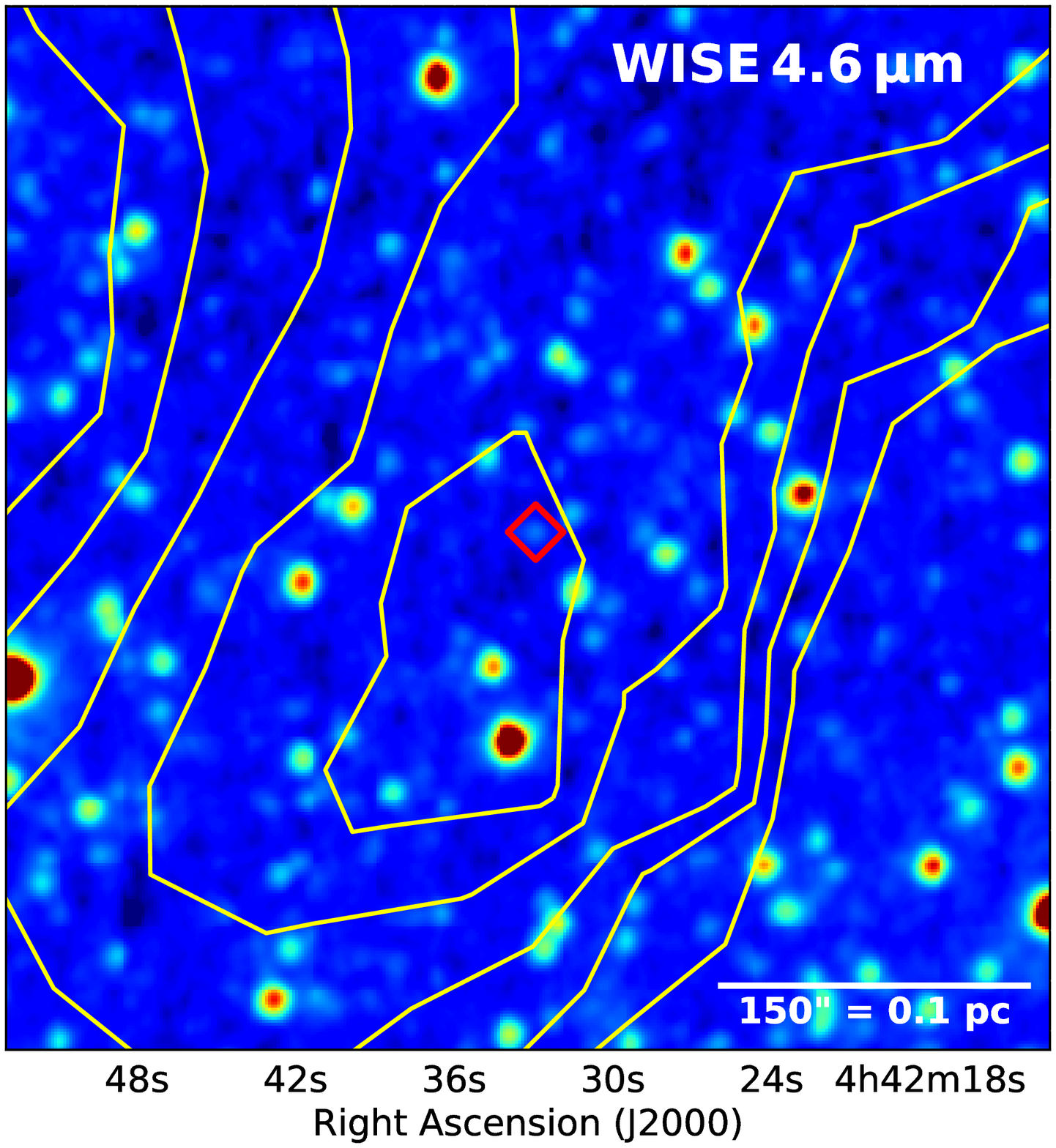}
\includegraphics[width=0.23\textwidth]{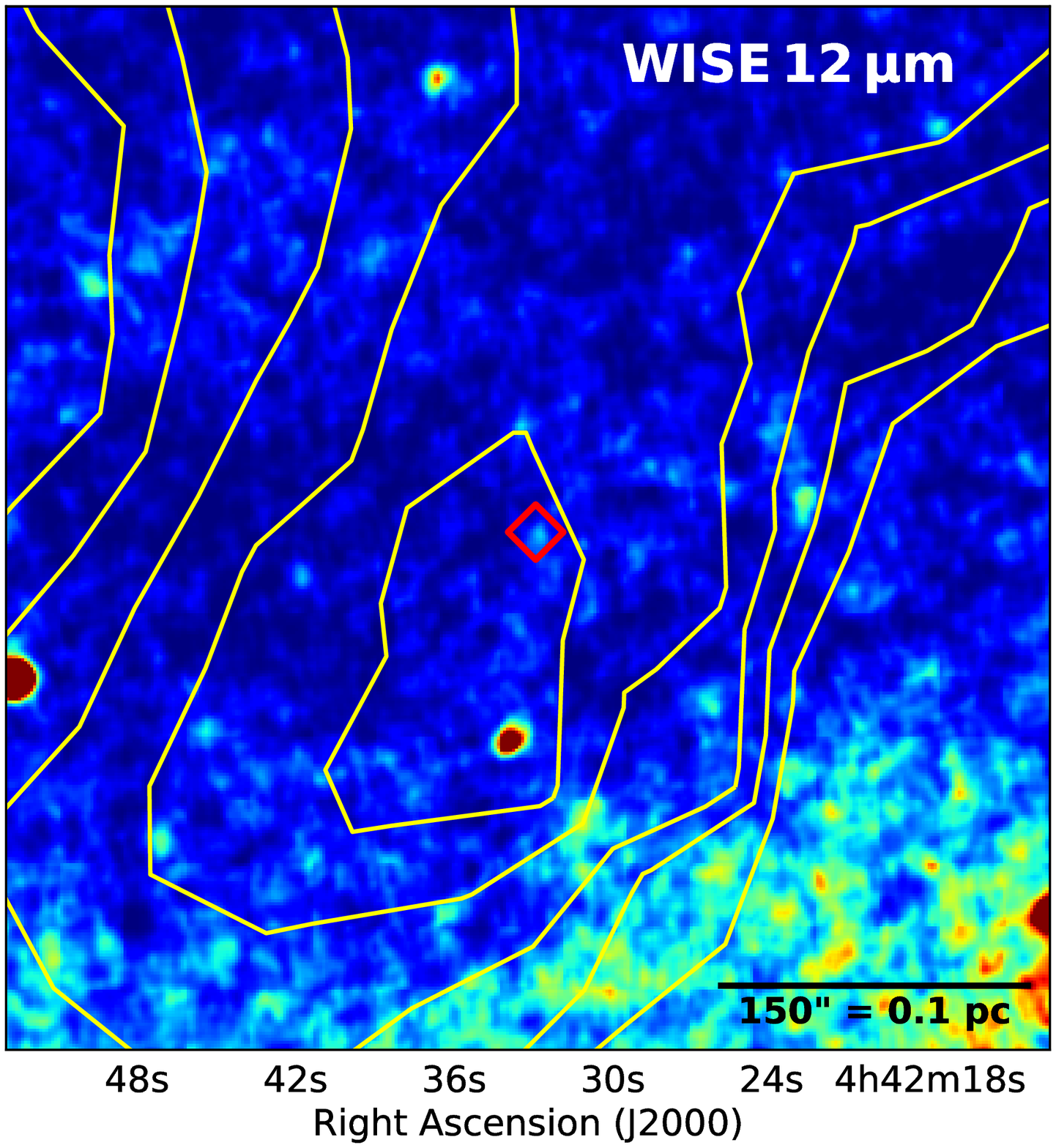}
\includegraphics[width=0.23\textwidth]{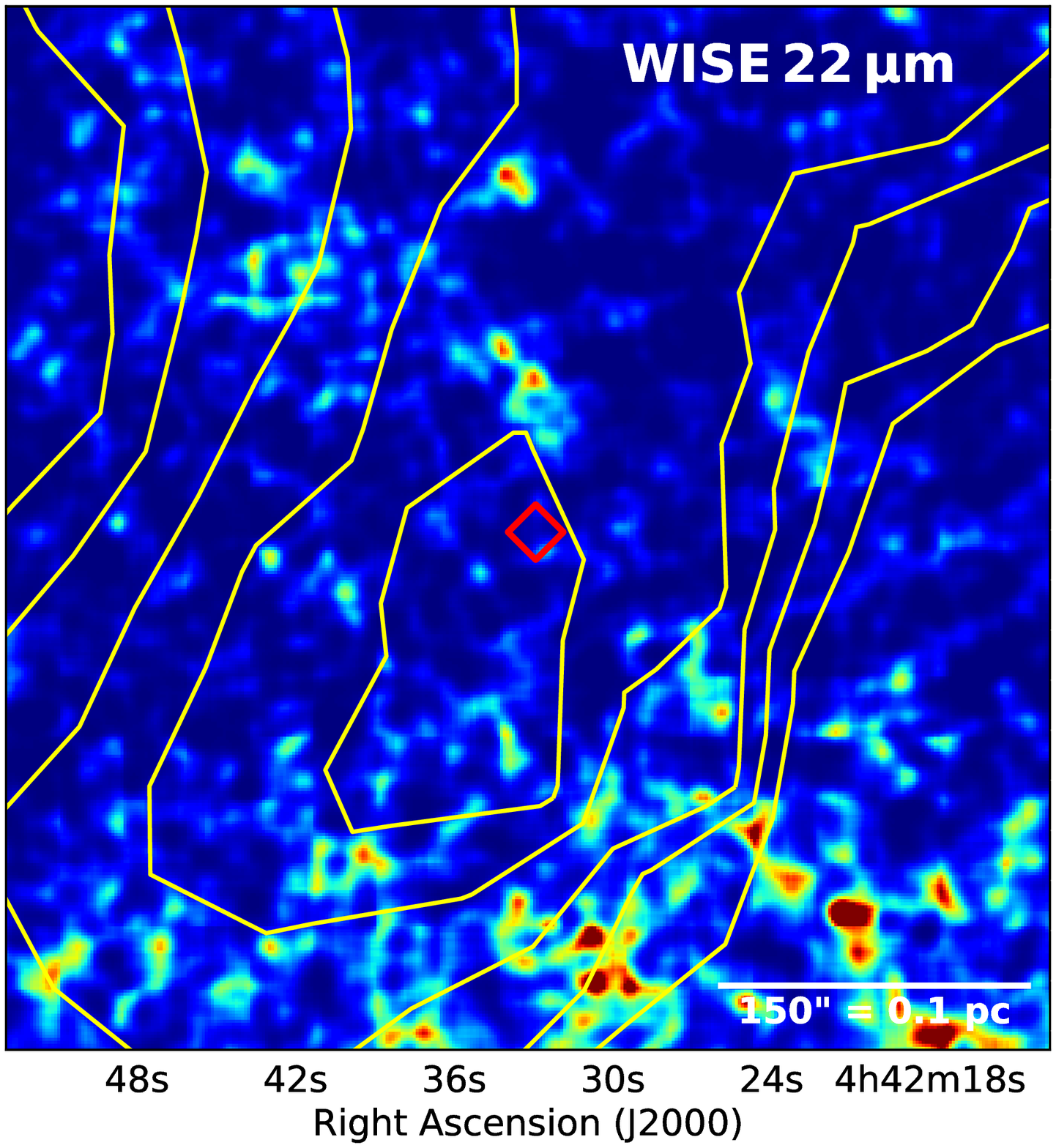}
\caption{\textit{WISE} W1--W4 images towards J044232.92 overlaid with contours of \textit{Planck} 550~$\mu$m dust continuum emission from G169.20-8.96. The W1 and W2 images are shown with square root scaling while the W3 and W4 images are shown with power-law scaling to enhance the colour contrast. The contours are as in Fig.~\ref{figure:planck}. Each panel is centred on the PGCC catalogue position of G169.20-8.96 and is $500\arcsec$ on a side. The red diamond symbol indicates the position of J044232.92. The scale bar corresponds to 0.1~pc at the distance of G169.20-8.96.}
\label{figure:images}
\end{center}
\end{figure*}

In Fig.~\ref{figure:planck}, we show a \textit{Planck} cutout image at 545~GHz (550~$\mu$m) towards G169.20-8.96. The \textit{WISE} images at all four IR wavelengths (W1--W4) towards J044232.92 are shown in Fig.~\ref{figure:images}. The \textit{WISE} photometric data of J044232.92 are given in Table~\ref{table:wise}, and selected properties of G169.20-8.96 are summarised in Table~\ref{table:planck}. We note that the clump masses and densities reported in the PGCC catalogue were derived assuming a dust-to-gas mass ratio of $R_{\rm dg}=1/100$ and a mean molecular weight of $\mu=2.33$ (\cite{planck2016}). However, the aforementioned value of $R_{\rm dg}$ refers to the canonical dust-to-hydrogen mass ratio, and the masses and densities listed in Table~\ref{table:planck} were scaled upwards by using a dust-to-gas mass ratio of $R_{\rm dg}=1/141$, which follows from the assumption that the chemical composition of the clump is similar to the solar mixture, where the hydrogen mass percentage is $X = 71\%$ and those of helium and metals are $Y = 27\%$ and $Z = 2\%$ (in which case the ratio of total gas mass (hydrogen, helium, and metals) to hydrogen gas mass is $(X + Y + Z)/X = 1.41$; e.g. \cite{anders1989}; 
\cite{kauffmann2008}). The aforementioned mean molecular weight refers to that per free particle, $\mu_{\rm p}$, and is valid for a $[{\rm He}]/[{\rm H}]$ ratio of 0.1 with no metals. However, we scaled the densities downwards by using a mean molecular weight per H$_2$ molecule, which in our case is $\mu_{\rm H_2}=2/X=2.82$ (\cite{kauffmann2008}, Appendix~A.1 therein).

\begin{table*}
\caption{Coordinates and photometric data of WISE~J044232.92+322734.9.}
\begin{minipage}{2\columnwidth}
\centering
\renewcommand{\footnoterule}{}
\label{table:wise}
\begin{tabular}{c c c c c c}
\hline\hline
$\alpha_{2000.0}$ & $\delta_{2000.0}$ & W1(3.4~$\mu$m) & W2(4.6~$\mu$m) & W3(12~$\mu$m) & W4(22~$\mu$m)\\

[h:m:s] & [$\degr$:$\arcmin$:$\arcsec$] & & & & \\
\hline 
04 42 32.92	& +32 27 34.9 & $15.601\pm0.051$ & $15.108\pm0.097$ & $12.017\pm0.345$ & $>8.254$ \\
 \ldots     &    \ldots   & $0.176\pm0.008$~mJy & $0.155^{+0.014}_{-0.013}$~mJy & $0.453^{+0.170}_{-0.123}$~mJy & $<3.723$~mJy \\   
\hline
\end{tabular} 
\tablefoot{The \textit{WISE} magnitudes given in the first row were taken from the AllWISE catalogue (the total in-band brightnesses). The W4 magnitude lower limit refers to the 95\% confidence brightness limit. The \textit{WISE} flux densities listed in the second row 
were computed from the magnitude values by applying the colour corrections under the assumption of a $S_{\nu}\propto \nu^{-2}$ power-law spectrum. Moreover, an additional W4 correction was applied to calculate the 22~$\mu$m flux density upper limit (see \cite{cutri2012}).}
\end{minipage} 
\end{table*}

\begin{table*}
\renewcommand{\footnoterule}{}
\caption{Properties of the dust clump PGCC~G169.20-8.96.}
\begin{minipage}{2\columnwidth}
\centering
\label{table:planck}
\begin{tabular}{c c c c c c c c c}
\hline\hline 
$\alpha_{2000.0}$ & $\delta_{2000.0}$ & $d$ & $D$ & $T_{\rm dust}$ & $\beta$ & $M$ & $n({\rm H_2})$ & $N({\rm H_2})$\\

[h:m:s] & [$\degr$:$\arcmin$:$\arcsec$] & [pc] & [pc] & [K] & & [M$_{\sun}$] & [$10^3$ cm$^{-3}$] & [$10^{21}$ cm$^{-2}$]\\ 
\hline 
04 42 33.22 & +32 27 36.80 & $140\pm15$ & $0.38\pm0.04$ & $12.2\pm2.0$ & $1.98\pm0.60$ & $3.4\pm2.0$ & $1.7\pm1.0$ & $1.3\pm0.7$\\
\hline
\end{tabular} 
\tablefoot{The coordinates refer to the PGCC catalogue position of G169.20-8.96. The other parameters in the table are the source distance ($d$), FWHM diameter of the clump ($D$), dust temperature derived from a modified blackbody fit to the clump SED ($T_{\rm dust}$), dust emissivity index corresponding to the SED fit ($\beta$), total mass of the clump ($M$), average H$_2$ number density of the clump ($n({\rm H_2})$), and H$_2$ column density of the clump. All the parameters were taken from the PGCC catalogue, but the mass and density values were scaled by using the present assumptions about the dust-to-gas mass ratio and mean molecular weight (see Sect.~2 for details).}
\end{minipage} 
\end{table*}

\section{Spectral energy distribution modelling of WISE~J044232.92+322734.9}

If J044232.92 is an external SFG, an obvious question is what is the redshift of the source. To determine the photometric redshift of 
J044232.92, we constructed its SED from the \textit{WISE} photometric data shown in Table~\ref{table:wise}, and fit the SED using 
an extension of the multiwavelength spectral modelling code {\tt MAGPHYS} (\cite{dacunha2008}, 2015) called {\tt MAGPHYS+photo-}$z$ (\cite{battisti2019})\footnote{All versions of {\tt MAGPHYS} are publicly available, and can be retrieved at 
\url{http://www.iap.fr/magphys/magphys/MAGPHYS.html}.}. Besides the photometric redshift, {\tt MAGPHYS+photo-}$z$ allows a simultaneous determination of the physical properties of the galaxy, such as its stellar and dust mass contents and the star formation rate (SFR). The uncertainties in the derived physical properties are propagated from the redshift uncertainty. The {\tt MAGPHYS} SED modelling is based 
on the assumption of a global energy balance between stellar and dust emissions, that is the ultraviolet (UV) and optical photons emitted by young stars are absorbed (and scattered) by dust grains in star-forming regions and more diffuse parts of the galactic interstellar medium, and the heated dust subsequently reradiates the absorbed energy in the IR. The dust heating is assumed to be predominantly caused by star formation activity, while the potential contribution from an AGN is not taken into account.

\begin{figure}[!htb]
\centering
\resizebox{\hsize}{!}{\includegraphics{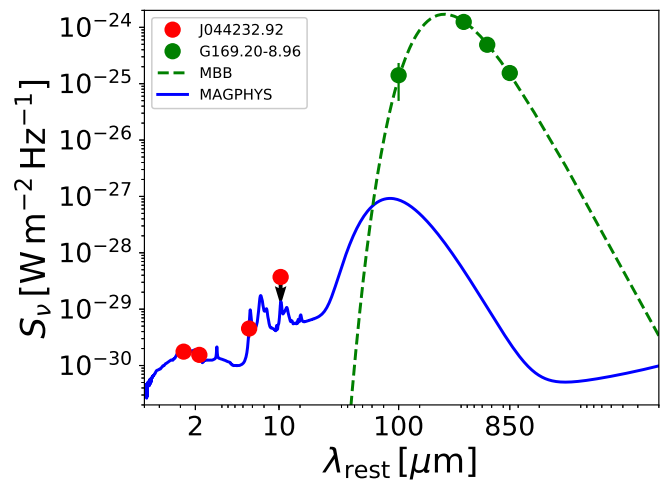}}
\caption{Best-fit rest-frame SED of J044232.92. The red points represent the observed photometric data points, and the one with a downwards pointing arrow marks the $\lambda_{\rm obs}=22$~$\mu$m upper flux density limit, which is taken into account in the fit. The blue line is the best-fit {\tt MAGPHYS+photo-}$z$ model SED (\cite{battisti2019}). For comparison, the far-IR to submillimetre SED of G169.20-8.96 is shown with green points indicating the \textit{IRAS} 100~$\mu$m and \textit{Planck} photometric data points and the green dashed line representing a single-temperature modified blackbody fit to those data (see text for details).}
\label{figure:sed}
\end{figure}

\begin{figure*}[!htb]
\begin{center}
\includegraphics[width=0.3\textwidth]{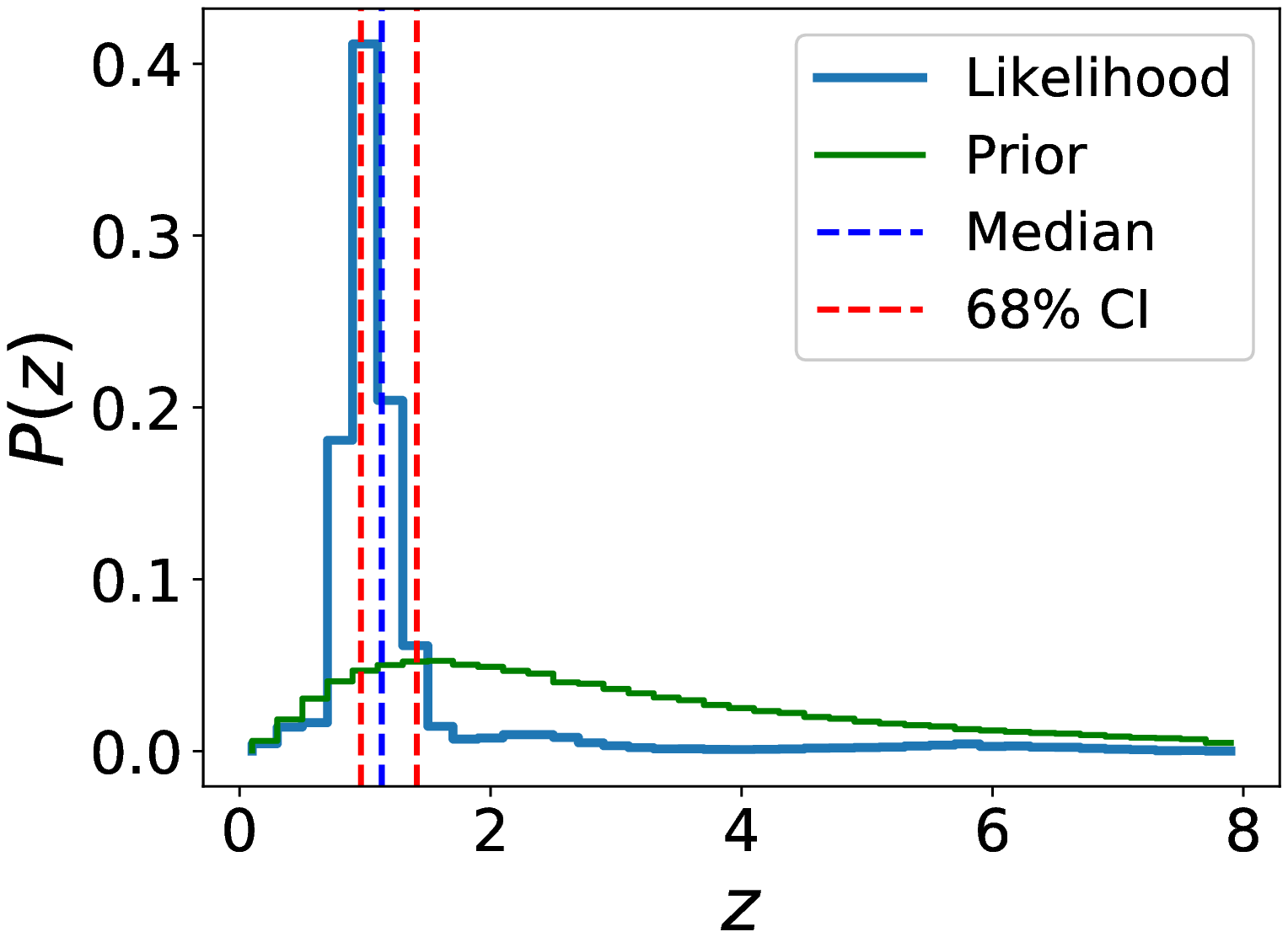}
\includegraphics[width=0.3\textwidth]{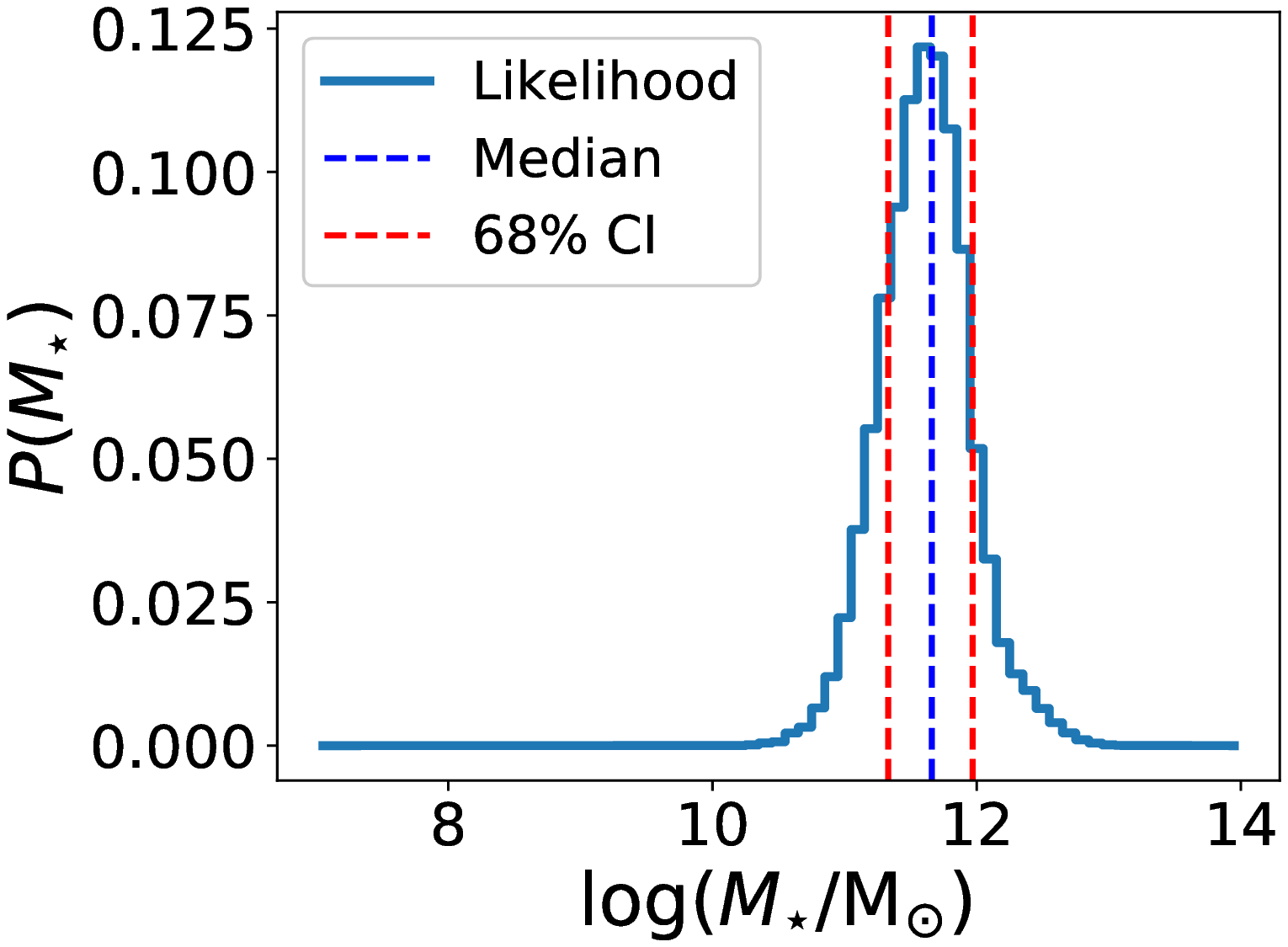}
\includegraphics[width=0.3\textwidth]{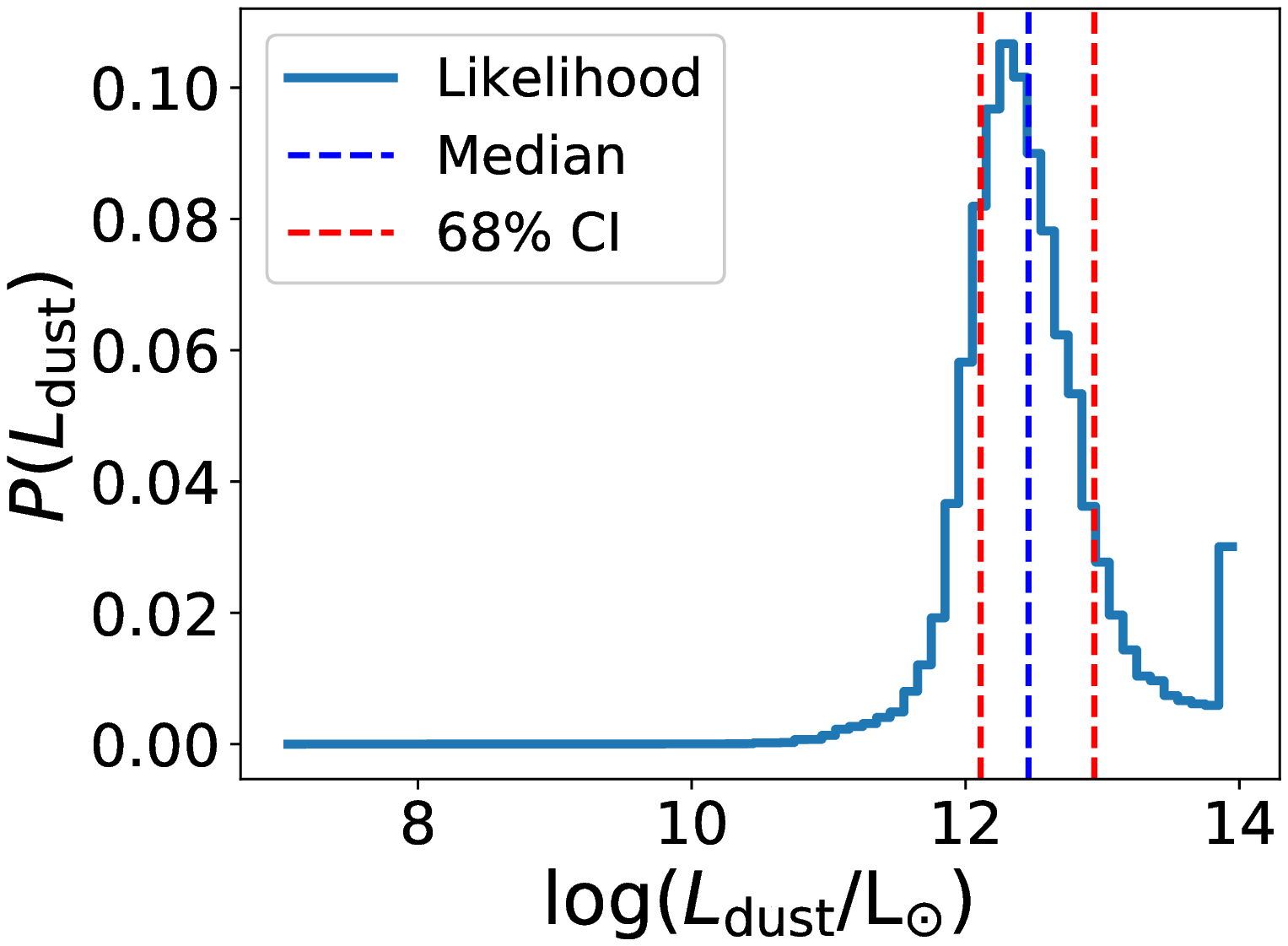}
\includegraphics[width=0.3\textwidth]{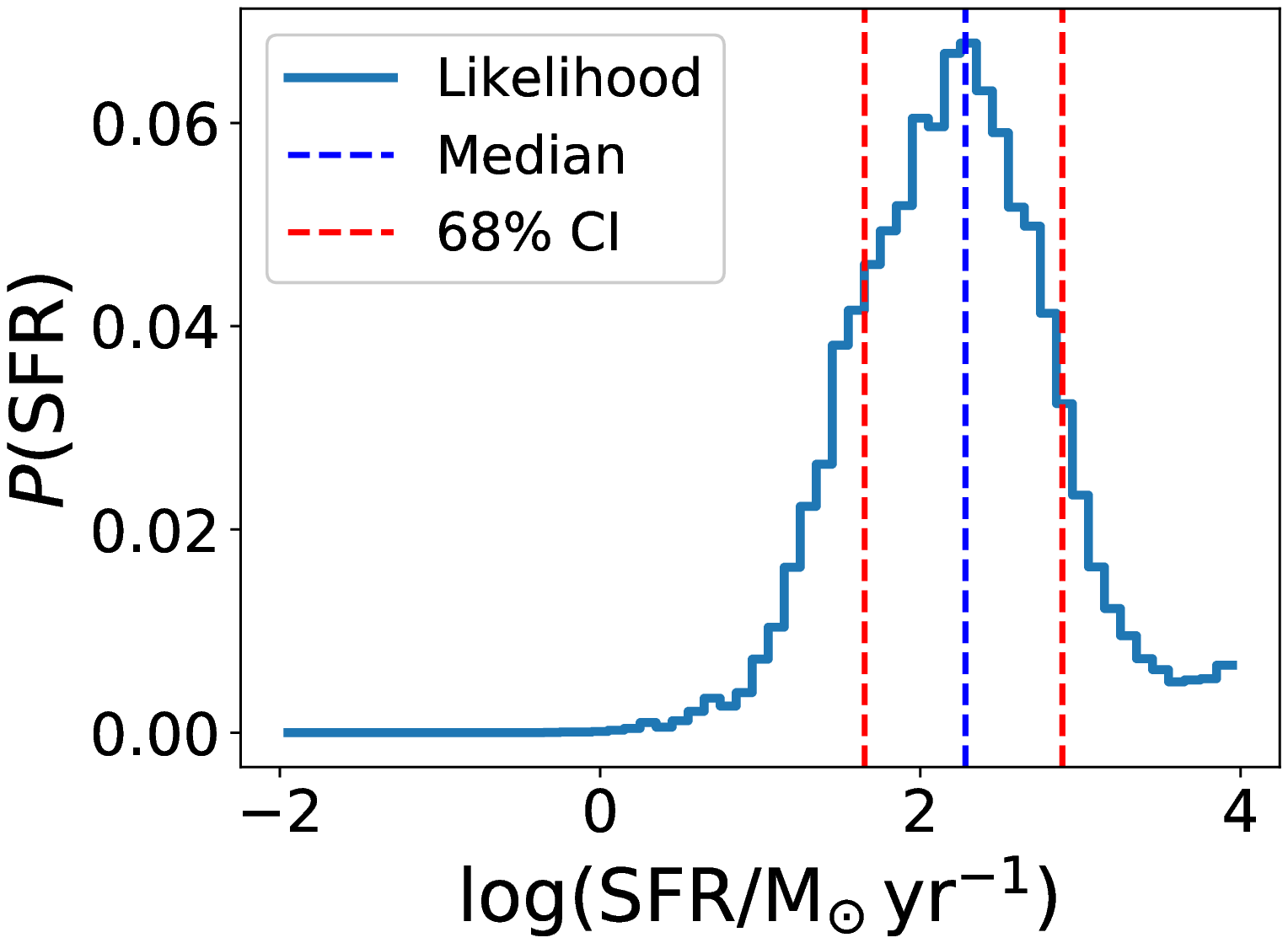}
\includegraphics[width=0.3\textwidth]{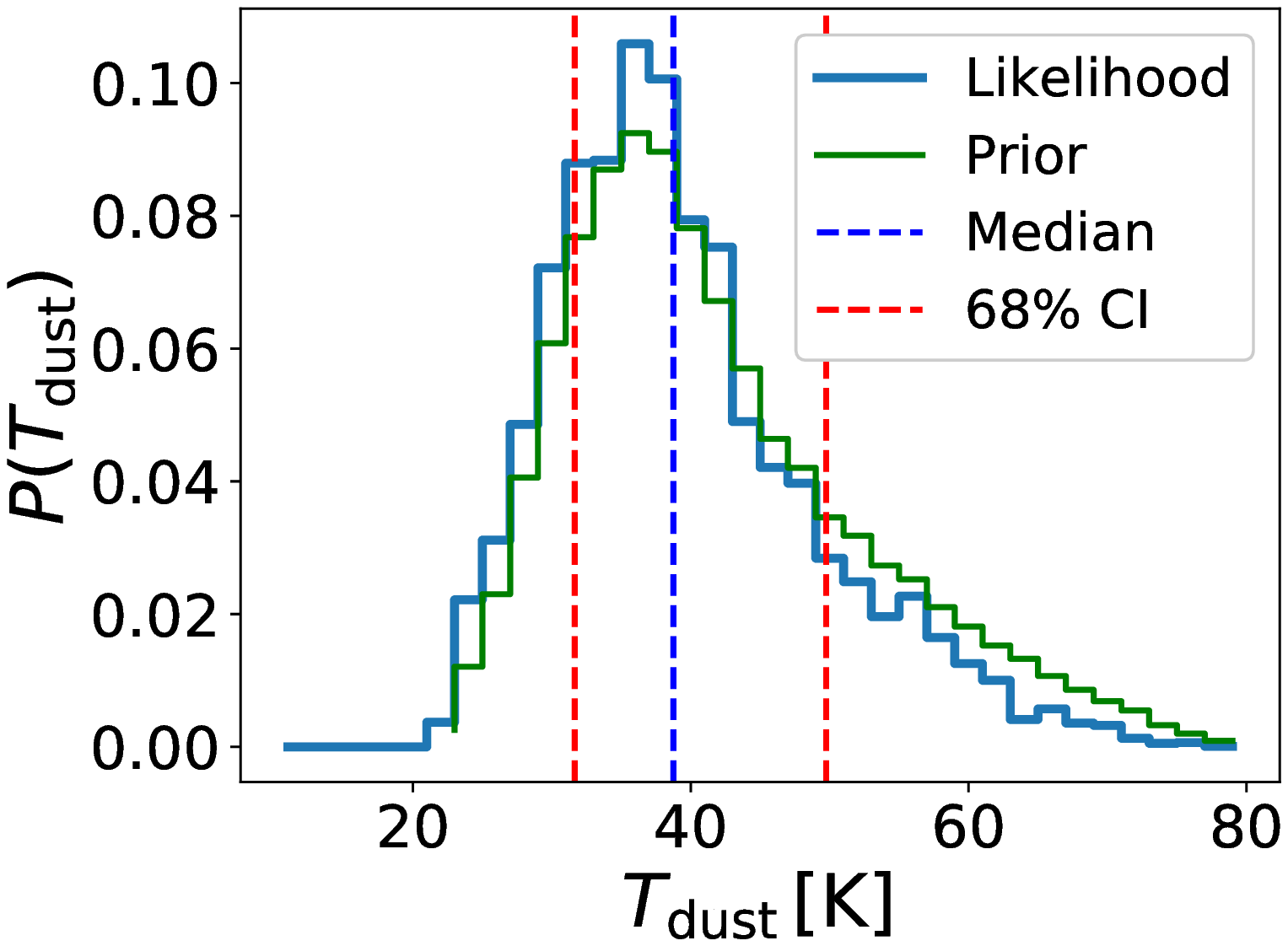}
\includegraphics[width=0.3\textwidth]{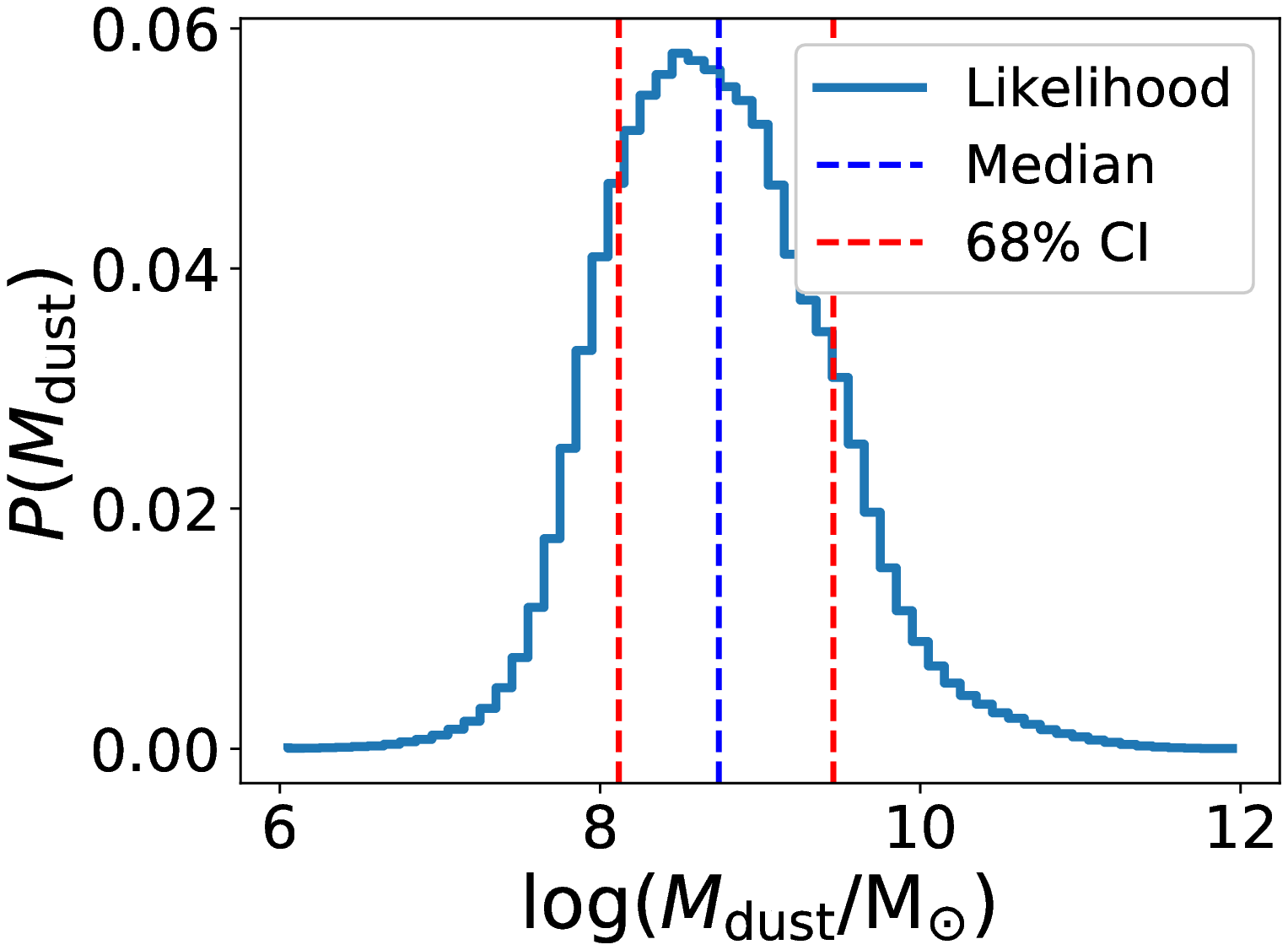}
\caption{Likelihood distributions of redshift, stellar mass, dust luminosity, SFR, dust temperature, and dust mass. For parameters with priors, the prior distribution is shown by a green line. In each panel, the blue dashed line shows the median value of the parameter with the 16th–-84th percentile range indicated by red dashed lines.}
\label{figure:likelihood}
\end{center}
\end{figure*}

The best-fit SED of J044232.92 is shown in Fig.~\ref{figure:sed} and the derived photo-$z$ and physical properties of J044232.92 are listed 
in Table~\ref{table:magphys}. The corresponding likelihood distributions are shown in Fig.~\ref{figure:likelihood}. Following da Cunha et al. (2015), Miettinen et al. (2017), and Battisti et al. (2019), the flux density upper limit (the observed-frame 22~$\mu$m data point) was taken into account in the SED fitting by setting the nominal value to zero, and using the upper limit value as the flux density uncertainty. As can be seen in Fig.~\ref{figure:sed}, the best-fit model is consistent with this upper limit value. We note that owing to the small number of data points in our SED (only three detections plus one left-censored data point), {\tt MAGPHYS} output a $\chi^2$ goodness-of-fit value (see Eqs.~(32) and (33) in da Cunha et al. (2008)) of zero for the best-fit model plotted in Fig.~\ref{figure:sed}.

For comparison with the SED of J044232.92, in Fig.~\ref{figure:sed} we also plot the SED of G169.20-8.96, which is constructed from the \textit{Planck} flux densities at 353, 545, and 857~GHz (i.e. at about 850, 550, and 350~$\mu$m; $S_{\rm 850\, \mu m}=15.5\pm2.3$~Jy, $S_{\rm 550\, \mu m}=49.1\pm7.0$~Jy, and $S_{\rm 350\, \mu m}=124.7\pm21.5$~Jy), and the 100~$\mu$m flux density of $S_{\rm 100\, \mu m}=14.1\pm9.2$~Jy measured with the \textit{Infrared Astronomical Satellite} (\textit{IRAS}; \cite{neugebauer1984}), which is also reported in the PGCC catalogue. The latter data were fitted with a single-temperature modified blackbody model (e.g. \cite{casey2012}; \cite{bianchi2013}), where we assumed that the dust emission is optically thin ($\tau \ll 1$). It was also assumed that the frequency-dependent dust opacity has the same form as that used by Planck Collaboration et al. (2016), that is the Beckwith et al. (1990) description $\kappa_{\nu}=0.1\times (\nu/1\, {\rm THz})^{\beta}$~cm$^2$~g$^{-1}$, where $\beta$ was fixed at 1.98 (see Table~\ref{table:planck}).

Besides the SFR derived from the {\tt MAGPHYS+photo-}$z$ SED modelling, the total IR luminosity in the rest-frame wavelength range 
of $\lambda_{\rm rest}=8$~$\mu$m--1~mm can be used to derive another estimate of the SFR. The standard ${\rm SFR}(L_{\rm IR})$ relationship from Kennicutt (1998; here scaled to a Chabrier (2003) IMF by dividing the original relationship by a factor of 1.7) is given by

\begin{equation}
\label{eqn:sfr_total_ir}
{\rm SFR}[{\rm M_{\sun}\,yr^{-1}}]=10^{-10}\times L_{\rm IR}[{\rm L_{\sun}}]\,.
\end{equation}
The Kennicutt (1998) relationship yields an SFR that refers to a stellar mass range of $0.1-100$~M$_{\sun}$ and which is averaged over the past 100~Myr (i.e. the same timescale to which our {\tt MAGPHYS} based SFR refers). The Kennicutt (1998) calibration is based on the assumptions of solar metallicity and an optically thick starburst region where UV photons are efficiently absorbed by dust grains, and hence reprocessed into IR photons. 

Equation~(\ref{eqn:sfr_total_ir}) yields an SFR of $281^{+569}_{-155}$~M$_{\sun}$~yr$^{-1}$, where the nominal value is 1.47 times higher than the one output by {\tt MAGPHYS+photo-}$z$. The difference arises from the fact that besides the newly forming stars, {\tt MAGPHYS} also allows for the heating of the dust by more evolved stellar populations, which is not taken into account in the Kennicutt (1998) calibration.

\begin{table*}
\renewcommand{\footnoterule}{}
\caption{Results of {\tt MAGPHYS+photo-}$z$ SED modelling of J044232.92.}
\begin{minipage}{2\columnwidth}
\centering
\label{table:magphys}
\begin{tabular}{c c c c c c c c c}
\hline\hline 
$z_{\rm phot}$ & $\log(M_{\star})$ & $\log(L_{\rm dust})$\tablefootmark{a} & ${\rm SFR}$\tablefootmark{b} & $T_{\rm dust}$ & $\log(M_{\rm dust}/{\rm M}_{\sun})$ \\

 & [M$_{\sun}$] & [L$_{\sun}$] & [${\rm M}_{\sun}$~yr$^{-1}$] & [K] & [M$_{\sun}$]\\ 
\hline 
$1.132^{+0.280}_{-0.165}$ & $11.66^{+0.31}_{-0.33}$ & $12.46^{+0.48}_{-0.35}$ & $191^{+580}_{-146}$ & $38.8^{+11.0}_{-7.1}$ & $8.74^{+0.72}_{-0.63}$ \\
\hline
\end{tabular} 
\tablefoot{The parameters given in the table are the photometric redshift ($z_{\rm phot}$), stellar mass ($M_{\star}$), dust luminosity in the rest-frame wavelength range of $\lambda_{\rm rest}=3$~$\mu$m--1~mm ($L_{\rm dust}$), SFR averaged over the past 100~Myr, luminosity-weighted dust temperature ($T_{\rm dust}$; see Eq.~(8) in \cite{dacunha2015}), and dust mass ($M_{\rm dust}$). The quoted uncertainties represent the 16th--84th percentile range of the likelihood distribution (68\% confidence interval).\tablefoottext{a}{The proportion of $L_{\rm dust}$ that is emitted in the far-IR range of $\lambda_{\rm rest}=42.5-122.5$~$\mu$m is $f_{\rm FIR}=0.895$, while that in the total-IR range of $\lambda_{\rm rest}=8$~$\mu$m--1~mm is $f_{\rm TIR}=0.976$.}\tablefoottext{b}{An SFR calculated using the ${\rm SFR}(L_{\rm IR})$ relationship of Kennicutt (1998) is $281^{+569}_{-155}$~${\rm M}_{\sun}$~yr$^{-1}$ (see Eq.~(\ref{eqn:sfr_total_ir})).}}
\end{minipage} 
\end{table*}

\section{Discussion}

\subsection{Nature of WISE~J044232.92+322734.9}

Owing to the small amount of data available for J044232.92, its SED is poorly constrained, and hence the physical properties of the source 
derived in Sect.~3 should be taken with caution. In particular, a reliable determination of the photo-$z$ would require data at multiple different wavelengths in the observed-frame optical and near-IR bands, which are typically only available for extragalactic survey fields such as the Cosmic Evolution Survey (COSMOS; \cite{scoville2007}). Moreover, the {\tt MAGPHYS+photo-}$z$ templates are optimised for $z>1$ SFGs (\cite{dacunha2015}; \cite{battisti2019}), which might not be appropriate for J044232.92 unless the source really is such an SFG. 

Da Cunha et al. (2015) and Miettinen et al. (2017) used the same high-$z$ {\tt MAGPHYS} model libraries as in the present study to derive 
the physical properties for their samples of SMGs. The stellar mass, dust luminosity, SFR, dust temperature, and dust mass derived for J044232.92 are 5.1, 0.8, 0.7, 0.9, and 1.0 times the nominal, full sample averages from da Cunha et al. (2015), and 4.2, 0.6, 0.4, 0.9, and 0.6 times the sample averages from Miettinen et al. (2017). We note that da Cunha et al. (2015) defined the SFR as the average over the past 10~Myr, and hence in the above comparison we used a similarly defined SFR, which is 184~${\rm M}_{\sun}$~yr$^{-1}$ for J044232.92. Also, the aforementioned SFR comparison with the Miettinen et al. (2017) value is based on the values derived from the Kennicutt (1998) relationship (Eq.~(\ref{eqn:sfr_total_ir})). Because the derived photo-$z$ of J044232.92, $z_{\rm phot}=1.132^{+0.280}_{-0.165}$, is lower than the typical redshifts of the SMGs in the samples of da Cunha et al. (2015; $z_{\rm median}=2.7$) and Miettinen et al. (2017; $z_{\rm median}=2.3$), the 4--5 times higher stellar mass of J044232.92 compared to the typical aforementioned SMGs could be an indication that the galaxy is more evolved and has assembled a higher stellar mass content, but the stellar mass of J044232.92 could also be overestimated owing to the lack of data in the UV, optical, and near-IR regime. Otherwise the physical properties of J044232.92 appear to be comparable to those of SMGs. Also, the derived total IR luminosity of J044232.92, $L_{\rm IR}=2.8^{+5.7}_{-1.5}\times10^{12}$~L$_{\sun}$, suggests that it is an ultraluminous IR galaxy (ULIRG; e.g. \cite{sanders1988}). 

To test how J044232.92 compares with the main sequence (MS) of SFGs, we used the Speagle et al. (2014) MS, which is defined as 

\begin{equation}
\begin{split}
\label{eqn:ms}
\log \left({\rm SFR}/{\rm M_{\sun}\,yr^{-1}}\right)_{\rm MS} & = (0.84-0.026\times \tau_{\rm univ})\log(M_{\star}/{\rm M}_{\sun})\\
& -(6.51-0.11\times \tau_{\rm univ})\,,
 \end{split}
\end{equation}
where $\tau_{\rm univ}$ is the age of the universe in Gyr ($\tau_{\rm univ}=5.259$~Gyr at the derived redshift of J044232.92). Using the nominal redshift, stellar mass, and SFR values derived with {\tt MAGPHYS+photo-}$z$, we obtain a starburstiness parameter, that is the distance from the MS, of $\Delta_{\rm MS}={\rm SFR}/{\rm SFR}_{\rm MS}=1.03$. If using the SFR derived from Eq.~(\ref{eqn:sfr_total_ir}), 
we obtained a value of $\Delta_{\rm MS}=1.51$. Both of these estimates of $\Delta_{\rm MS}$ suggest that J044232.92 is a MS galaxy (see e.g. \cite{miettinen2017} and references therein for different MS offset criteria; Sect.~4.1.1 therein).  

\subsection{Nature of PGCC~G169.20-8.96}

The coordinates and distance of G169.20-8.96 suggest that it is a member of the Taurus–Auriga molecular cloud complex at $d=140$~pc (e.g. \cite{kenyon1990}; see also \cite{gudel2007} and \cite{kenyon2008} for reviews). Unfortunately, G169.20-8.96 lies outside the regions of panchromatic observations of Taurus (see e.g. Fig.~1 in \cite{gudel2007}), for example those in the optical with the MegaCam on the Canada-France-Hawaii Telescope (\cite{boulade2003}) and in the near to mid-IR with \textit{Spitzer} (\cite{werner2004}). Moreover, G169.20-8.96 is outside of the \textit{Herschel} (\cite{pilbratt2010})\footnote{\textit{Herschel} is an ESA space observatory with science instruments provided by European-led Principal Investigator consortia and with important participation from NASA.} fields observed as part of the Gould Belt Survey (GBS; \cite{andre2010})\footnote{\url{http://gouldbelt-herschel.cea.fr}} as shown in Fig.~1 in Kirk et al. (2013). 

The clump is cold, $T_{\rm dust}=12.2\pm2.0$~K, and has a mass of $3.4\pm2.0$~M$_{\sun}$ within a FWHM diameter of $0.38\pm0.04$~pc. Hence, G169.20-8.96 is a potential low-mass star-forming object. As seen in Fig.~\ref{figure:planck}, G169.20-8.96 appears to be part of a system of three clumps that are blended in the \textit{Planck} 550~$\mu$m map. The beam-averaged H$_2$ column density of G169.20-8.96 is fairly low, $N({\rm H_2})=(1.3\pm0.7)\times10^{21}$~cm$^{-2}$, but higher resolution observations would undoubtedly reveal denser substructures within the clump. 

Because the distance to G169.20-8.96 is only 140~pc, its angular size on the sky is large, $9\farcm33$ in FWHM diameter. This, together with its high absolute Galactic latitude of $\vert b \vert=8\fdg96$ increase the probability for an extragalactic object seen along the line-of-sight towards the clump.

\subsection{Implications for studies of dust-obscured star-forming galaxies and Galactic molecular cloud clumps}

Survey studies of high-redshift, dusty SFGs, especially the SMGs, often start 
by imaging an extragalactic field with a single-dish telescope at a (sub-)millimetre wavelength 
band (e.g. \cite{scott2008}; \cite{weiss2009}; \cite{aretxaga2011}). A common, subsequent step 
is to try to identify the multiwavelength counterparts 
of the identified (sub-)millimetre sources in order to derive their panchromatic 
SEDs and photometric redshifts. If, however, some of the originally uncovered (sub-)millimetre sources 
belong to the Galaxy, one could mistakenly interprete some galaxy to 
be an SMG although it would not be one. For example, the derivation 
of the photometric redshift of the counterpart galaxy could be based on the 
optical to near-IR data alone, and hence the presence of a Galactic 
contaminant could be missed. 

As described above, a Galactic dust clump such as G169.20-8.96 in the Taurus–Auriga molecular cloud complex could erroneously be attributed to the dust emission of a dusty SFG, most notably an SMG. However, G169.20-8.96 is a much brighter submillimetre source than SMGs, and for example its 353~GHz or 850~$\mu$m flux density is reported to be $S_{\rm 850\, \mu m}=15.5\pm2.3$~Jy in the PGCC catalogue (Sect.~3). For comparison, the brightest Large APEX BOlometer CAmera (LABOCA; \cite{siringo2009}) 870~$\mu$m selected SMG in the Extended Chandra Deep Field South (ECDFS) has an 870~$\mu$m flux density of 14.5~mJy (\cite{weiss2009}), which is over 970 times lower than that of G169.20-8.96 ($S_{\rm 870\, \mu m}=14.1\pm1.9$~Jy, which was calculated from $S_{\rm 850\, \mu m}$ assuming that $S_{\nu}\propto \nu^{2+\beta}$ and using the $\beta$ value from Table~\ref{table:planck}). However, a direct comparison between the \textit{Planck} and LABOCA flux densities is not totally fair because the \textit{Planck} beam at 850~$\mu$m is over 14 times larger than that of LABOCA, and the spatial filtering by LABOCA makes it less sensitive to diffuse, large-scale emission seen by \textit{Planck}. Nevertheless, a Galactic dust clump would be fairly easy to distinguish from a high-$z$ SMG. A suspiciously bright (sub-)millimetre source could also be observed in spectral line emission (or absorption) to determine whether its radial velocity (and the corresponding kinematic distance) places it within the Galaxy and, hence, whether it is physically unrelated with the putative counterpart.

Moreover, although the Galactic latitude of G169.20-8.96 is low ($b=-8\fdg96$), extragalactic survey fields have much higher absolute Galactic latitudes, $\vert b \vert$, where the occurrence of Galactic dust cores and clumps is lower than closer to the Galactic plane where molecular clouds are concentrated in (e.g. \cite{dame2001}). For example, $b=-54\fdg44$ for the ECDFS (\cite{giacconi2001}, 
\cite{giacconi2002}) and $b=42\fdg12$ for COSMOS. Nevertheless, nearby dust clumps such as G169.20-8.96 might still be a concern.

On the other hand, an SFG, which appears as a mid-IR point source, could mistakenly be interpreted to be a 
YSO embedded in a Galactic molecular cloud clump or core seen in projection towards the SFG. In this case, the clump or core could be 
misclassifed as protostellar although it could really be starless (and prestellar if it is also gravitationally bound). Such misclassification could 
then propagate to other, subsequent analyses, such as the duration of the starless 
phase of core evolution, which is sometimes derived from the relative 
numbers of starless and protostellar cores (the so-called statistical lifetime estimator; 
e.g. \cite{jorgensen2007}; \cite{enoch2008}; \cite{evans2009}).

\section{Summary and conclusions}

In the present paper, we have presented a potential example of a case where an external galaxy is seen in projection towards a dust clump in the Milky Way galaxy. The background extragalactic source is WISE~J044232.92+322734.9, which was identified through IR observations with \textit{WISE}, while dust clump in question is PGCC~G169.20-8.96 that was uncovered by submillimetre observations with the \textit{Planck} satellite. The dust clump lies at a distance of 140~pc, and is likely associated with the Taurus-Auriga complex. 

We used the {\tt MAGPHYS+photo-}$z$ code to derive a photometric redshift estimate for J044232.92 and its basic physical properties. The photo-$z$ was derived to be $z_{\rm phot}=1.132^{+0.280}_{-0.165}$, while the stellar mass, IR (8--1\,000~$\mu$m) luminosity, SFR, dust temperature, and dust mass were derived to be $M_{\star}=4.6^{+4.7}_{-2.5}\times10^{11}$~M$_{\sun}$, $L_{\rm IR}=2.8^{+5.7}_{-1.5}\times10^{12}$~L$_{\sun}$, $191^{+580}_{-146}$~${\rm M}_{\sun}$~yr$^{-1}$ ($281^{+569}_{-155}$~${\rm M}_{\sun}$~yr$^{-1}$ from $L_{\rm IR}$), 
$T_{\rm dust}=38.8^{+11.0}_{-7.1}$~K, and $M_{\rm dust}=5.5^{+23.3}_{-4.2}\times10^{8}$~M$_{\sun}$. The derived properties of J044232.92 suggest that it is a ULIRG that lies in the main sequence of SFGs. Its physical properties are also comparable to those of SMGs in the ECDFS and COSMOS fields. However, to our knowledge, J044232.92 has only been observed with \textit{WISE}, and hence its SED could not be well constrained on the basis of the currently available data (detections in the \textit{WISE} bands W1--W3 (3.4~$\mu$m--12~$\mu$m) and a non-detection in the W4 (22~$\mu$m) band). Therefore, the derived redshift and physical properties of J044232.92 should be considered as much more uncertain than the quoted formal uncertainties. 

(Sub-)millimetre dust continuum and radio continuum imaging at high angular resolution would help to test the hypothesis that J044232.92 really is an SFG as suggested by its \textit{WISE} IR colours, and whether it could belong to the SMG population at a redshift of about 1. High-resolution interferometric imaging would allow us to pinpoint the exact location and spatial extent of the potential dust-emitting region of J044232.92. For example, the present SED model prediction is that the observed-frame 850~$\mu$m flux density of J044232.92 is $\sim3$~mJy, which would be only about 0.2 per mille of the \textit{Planck} 850~$\mu$m flux density of G169.20-8.96. Follow-up observations of J044232.92 would be useful because they could reveal the presence of an SMG outside extragalactic survey fields and which could be a rare type of a projection of physically unassociated sources.

\begin{acknowledgements}

I would like to thank the anonymous referee for providing comments and suggestions. This research has made use of NASA's Astrophysics Data System Bibliographic Services, the NASA/IPAC Infrared Science Archive, which is operated by the Jet Propulsion Laboratory, California Institute of Technology, under contract with the National Aeronautics and Space Administration, and {\tt Astropy}\footnote{\url{http://www.astropy.org}}, a community-developed core Python package for Astronomy (\cite{astropy2013}, \cite{astropy2018}). This publication makes use of data products from the \textit{Wide-field Infrared Survey Explorer}, which is a joint project of the University of California, Los Angeles, and the Jet Propulsion Laboratory/California Institute of Technology, and NEOWISE, which is a project of the Jet Propulsion Laboratory/California Institute of Technology. \textit{WISE} and NEOWISE are funded by the National Aeronautics and Space Administration.

\end{acknowledgements}

\end{document}